\documentclass[useAMS,usenatbib]{mn2e}

\usepackage{amsmath}
\usepackage{rotating}
\usepackage{color}
\usepackage{times}
\usepackage{slashbox}
\usepackage{graphicx, subfigure}
\usepackage{multirow}
\usepackage{array}
\usepackage{epsfig}
\usepackage{epstopdf}
\usepackage{float}
\usepackage{gensymb}
\usepackage{enumerate}

\newif\ifpdf
\ifx\pdfoutput\undefined
   \pdffalse
\else
   \pdfoutput=1
   \pdftrue
\fi
\ifpdf
   \usepackage{graphicx}
   \usepackage{epstopdf}
\else
   \usepackage{graphicx}
\fi

\newcommand{\kpc}{{\rm\,kpc}}
\newcommand{\kms}{{\rm\,km\,s^{-1} }}
\newcommand{\msun}{{\rm\,M_\odot }}

\title[Stability of Satellite Planes in M31 II]{Stability of Satellite Planes in M31 II: Effects of the Dark Subhalo Population}
\author[Nuwanthika Fernando.]{Nuwanthika Fernando$^{1}$\thanks{E-mail:
nuwanthika.fernando@sydney.edu.au}, Veronica Arias$^{2}$, Geraint F. Lewis$^{1}$, Rodrigo A. Ibata$^{3}$,\and Chris Power$^4$\\  
$^{1}$Sydney Institute for Astronomy, School of Physics A28, The University of Sydney, NSW 2006, Australia\\
$^{2}$Departamento de Fisica, Universidad de los Andes, Cra. 1 No. 18A-10, Edificio Ip, Bogota, Colombia\\
$^{3}$Observatoire de Strasbourg, 11, rue de l'Universit\'e, F-67000, Strasbourg, France\\
$^{4}$ICRAR, University of Western Australia, 35 Stirling Highway, Crawley, Western Australia 6009, Australia
\\} 
\begin{document}

\date{Accepted 2017 xxx. Received 2017xxx; in original form 2017 June Date}

\pagerange{\pageref{firstpage}--\pageref{lastpage}} \pubyear{2016}

\maketitle

\label{firstpage}

\begin{abstract}

The planar arrangement of nearly half the satellite galaxies of M31 has been a source of mystery and speculation since it was discovered. With a growing number of other host galaxies showing these satellite galaxy planes, their stability and longevity have become central to the debate on whether the presence of satellite planes are a natural consequence of prevailing cosmological models, or represent a challenge. Given the dependence of their stability on host halo shape, we look into how a galaxy plane's dark matter environment influences its longevity. An increased number of dark matter subhalos results in increased interactions that hasten the deterioration of an already-formed plane of satellite galaxies in spherical dark halos. The role of total dark matter mass fraction held in subhalos in dispersing a plane of galaxies present non trivial effects on plane longevity as well. But any misalignments of plane inclines to major axes of flattened dark matter halos lead to their lifetimes being reduced to $\leq$3 Gyrs. Distributing $\geq$ 40\% of total dark mass in subhalos in the overall dark matter distribution results in a plane of satellite galaxies that is prone to change through the 5 Gyr integration time period. 

\end{abstract}


\section{Introduction}

The first hints of anisotropic distributions in Milky Way's (MW) satellite galaxies were seen nearly four decades ago, as \citet{Lynden-Bell1976}\ and \citet{1976RGOB..182..241K}\ discovered several satellites aligned with the orbits of the Magellanic Clouds. The planar nature of this arrangement became clear with surveys that offered more detailed analysis in later years \citep[e.g.][]{Lynden-Bell1982,Pawlowski2014} and its current members include 11 of the brightest satellites and streams of the MW. Another instance of such anisotropic distribution was discovered by the Pan-Andromeda Archaeological Survey (PAndAS), which placed our nearest neighbour Andromeda (M31) and its surroundings under scrutiny \citep{Ibata2013, 2013ApJ...766..120C}. It was named the Vast Thin Plane of Satellites (VTPoS) and earned it by hosting 15 of the known 30 satellites galaxies in a planar formation extending over a 200\kpc\ radial distance, and only 12.6$\pm$0.6\kpc\ in thickness. Adding to these impressive dimensions, the VTPoS shows signs of co-rotation between 13 of its members and the probability of a similar 'visual' plane (a superficial planar arrangement in our line of sight) occurring in a randomly distributed satellite group approached 0.13\% \citep{Ibata2013}. Recent doubts about the validity of the the MW's Vast Polar Structure (VPS) of satellites and streams \citep{2017arXiv170200485M,2017arXiv170200497M} have been countered by \citet{2017arXiv170206143P} as well. Planar arrangements of galaxies were found further out into the Local Volume, when \citet{Tully2015} gave evidence for two parallel systems in the group Centaurus (although recent work of \citet{2016A&A...595A.119M} show only one plane of significance when newly discovered satellite galaxies in Centaurus were included in calculations). 
 
As satellite galaxy planes were observed in both large galaxies in our Local Group, they began to take a significant place in local small-scale structure formation. If these planar formations are to be considered long-term kinematic structures, their longevity and stability need to be investigated. From a variety of $\Lambda$CDM based cosmological simulations, we find that the dark matter of MW and M31-sized galaxies is not smoothly distributed but is rather clumpy instead, where an assortment of smaller completely dark subhalos populate a dark matter environment \citep{Boylan-Kolchin2009, Sawala2016}. As these dark subhalos would alter the nature of gravitational interactions of the satellites with the overall dark matter distribution, a crowded or sparse dark subhalo environment may affect the longevity of a plane of satellite galaxies.  

We previously examined the stability of satellite planes in \citet{Fernando2016}, referred hereafter as Paper I, for formations belonging to host galaxies similar to M31 in a MW+M31-like system. Its results showed that the most stable planes are those that are dynamically cold and aligned with the host's dark matter halo axes. This paper extends our exploration of the stability of planes to investigate the aspects of the immediate environment of these structures. Our planar arrangements of satellite galaxies are evolved in a variety of surroundings of dark subhalos that simulations show to occupy the virial radius of a host. Following a background on dark subhalos and satellite plane stability in Sec.~\ref{sec:Bgd}, Sec.~\ref{sec:NumMod} paints a picture of the models and methods used in the paper. Sec.~\ref{Stb_DH} looks at a satellite's probability of being on its initial plane for the given integration time under the influence of factors like incline and halo shape, to test if their effects are augmented or diminished by the inclusion of dark subhalos. Sec~\ref{sec:Disc} will analyse the results of the tests by comparing observations and simulations to provide predictions or limitations on these properties. 

\section{Background}\label{sec:Bgd}
Large-scale cosmological simulations, consisting of billions of particles, have proven to be useful laboratories to test theories on the formation and evolution of galaxies. Given the complexity of baryonic physics, the majority of such simulations have been dark-matter-only, representing $\Lambda$ Cold Dark Matter ($\Lambda$CDM) cosmology (e.g. \citet{Boylan-Kolchin2009} with Millennium II). Whilst the picture that emerges from $\Lambda$CDM numerical simulations agrees very well with large-scale observations, they apparently disagree when describing small-scale structure. In simulations, MW-sized halos would contain many thousands of subhalos that are expected to host satellite galaxies, which leads to the `missing satellite problem'- where the predicted number does not match the observed number of satellite galaxies in the vicinity of the MW and M31 \citep{1999ApJ...522...82K, 1999ApJ...524L..19M}. Suggested solutions include alternative cosmologies, such as  Warm Dark Matter \citep{2017MNRAS.464.4520B} which cuts off the power spectrum at low masses. But many feel that the most straightforward interpretation of this absence in $\Lambda$CDM is that these subhalos were stripped of their baryonic content due to astrophysical processes in the early universe. Another potential problem is `Too-Big-to-Fail', where the distribution of subhalo masses show an overabundance of large-mass satellites compared to those observed in the local universe \citep{2010ApJ...717.1043B, 2011MNRAS.415L..40B}. Again, astrophysical processes of the early universe are offered as the most likely potential solution. %

More recently, attention has turned to the distribution of satellite galaxies, both theoretically and in observations, as planar formations were observed in the MW and M31. Though \citet{Ibata2014} finds an overabundance of co-rotating pairs of satellite galaxies in the SDSS, the distribution of satellite galaxies in $\Lambda$CDM simulations show roughly isotropic distributions. By relaxing the search criteria to accomodate diverse numbers of members (or fraction of memeber/non-member satellite galaxies) and plane thicknesses, these arrangements stop being as uncommon as previously thought in simulations \citep{Cautun2015a}. However, large and small-scale simulations are as not unsuccessful in accurately recreating the extremely thin plane of M31, with a similar fraction of members \citep{Buck2015, 2015MNRAS.453.3839P}. While there is hope that the inclusion of baryons \citet{Sawala2016} and higher resolution numerical simulations would provide a natural solution to the presence of satellite planes in $\Lambda$CDM cosmology, \citet{2015ApJ...815...19P}, concludes that this problem is unlikely to be solved solely with the inclusion of baryonic astrophysics. Recent work by \citet{2017MNRAS.466.3119A} reveal that MW-sized simulations that include baryons give a variation of identifiable satellite planes, but when corotation and thickness of M31 is considered, planes similar to the VTPoS were not detected. 

The arrangement of satellite galaxy planes currently observed within the Local Group is rather peculiar \citep{2013MNRAS.435.1928P, 2015MNRAS.452.1052L}. Our MW's VPS is formed orthogonal to the Galactic disk \citep{Pawlowski2012b}, the neighbouring VTPoS is seen with its normal at $\sim90^{\circ}$ a with the MW-M31 vector.  \citet{doi:10.1093/mnras/stx286} suggests that the possible transient nature of satellite planes is why such formations are sparse in $\Lambda$CDM simulations, but does not address why we see two such structures in the Local Universe today.  

Despite the cumulation of new simulations and observations, questions about the survivability and lifespan of satellite galaxy planes still remain unanswered.  In Paper I we explored how the longevity of satellite planes is affected by intrinsic properties (e.g. mass and velocity perpendicular to the plane) and host dark matter halos (e.g. shape and plane alignment to host halo axes). It's results lead us to  conclusions that agreed with the general consensus of similar papers \citep[e.g.][]{Bowden2013}. Maintaining a high probability of survival for 5 Gyrs required planes to be dynamically cold and to originate with perpendicular velocities that less than 30 $\kms$ perpendicular to the plane. Higher perpendicular velocities reduced the probability of seeing a satellite (initially set on a plane) to values that are similar to finding a superficial plane in a random distribution of satellites, in less than 3 Gyrs. For kinematically stable planes, those aligned with a major axis of the host's dark halo (in non-spherical halo forms) maintain a high probability of survival through the integration period, compared to those that are misaligned. Although tidal streams are expected to precess through the years in flattened halos \citep{Ibata2001}, our results presented little or no evidence that flat halos caused `precession' in satellite planes, over a 5 Gyr period. 

However, the tests of Paper I were conducted in a smooth dark matter halo, void of the hundreds, if not thousands of possible dark subhalos populating the dark halo locale of M31 or MW-like  galaxies. Therefore, to better understand the behaviour of satellite planes in this dark substructure, it is necessary to introduce dark subhalos in further simulations.  

\section{Numerical Model}\label{sec:NumMod}

Our numerical model replicates the model used in Paper I- which created a plane of satellite galaxies within a combination of rigid potentials (halo, disk and bulge) that represented M31 \citep{2006MNRAS.366..996G,Arias2016}. 

The dark matter halo is described by a Navarro-Frenk-White (NFW) potential \citep{1997ApJ...490..493N};
\begin{equation}
\label{EqNFW}
\Phi_{\rm{halo}}(r)=-4{\pi}G\delta_{\rm{c}}\rho_{\rm{c}}{r_{\rm{h}}}^2\left(\frac{r_{\rm{h}}}{r}\right)\log \left(\frac{r+r_{\rm{h}}}{r_{\rm{h}}}\right),
\end{equation}   
where $r_{\rm{h}}$ is the scale radius, the present day critical density is $\rho_{\rm{c}}=277.7\,h^2\msun\,{\rm{kpc^{-3}}}$, $h=0.71$ in units of 100 $\kms\rm{Mpc^{-1}}$ \citep{2006MNRAS.366..996G}  and $\delta_{\rm{c}}$ is a dimensionless density parameter.

The halo mass values calculated from observations include the masses of these subhalos found inside the virial radius of the host halo. Therefore, when adding dark halos to the environment of the M31 potential we wanted to maintain the total host halo mass. For the work considered in this paper, we reduce the diffuse M31 halo mass by varying the `$halo\ mass\ fraction$', $h_{m}$, and by replacing $\delta_{\rm{c}}$ in Eq.~\ref{EqNFW} with $\delta_{\rm{Nc}}$ defined as: 
\begin{equation}
\label{EqH}
    \delta_{\rm{Nc}} = \delta_{\rm{c}}h_{m} .  
\end{equation}

The dark matter halos of galaxies the size of the MW and M31 are not seen in cosmological simulations to be perfectly spherical \citep{2009MNRAS.398.1150B, 2008MNRAS.391.1940M}. To further study the the effects of non-spherical halos, we modified our initial numerical model to introduce a $flatness$ = $q$ to the $z$ axis of the standard NFW equation (Eq.~\ref{EqNFW})- flattening the halo shape by reducing $q$ value. We replace $r^2$ of $\Phi_{\rm{halo}}(r)$ with: %
\begin{equation}
\label{EqFl}
	r^2=x^2 + y^2 + (z/q)^2 .
\end{equation}

The disk component of the potential remains
\begin{equation}
\Phi_{\rm{disk}}(r)=-2{\pi}G\Sigma_{0}{r^{2}_{\rm{disk}}}\left[\frac{1-\exp^{-r/r_{\rm{disk}}}}{r}\right]
\end{equation}   
and the bulge component follows \citet{1990ApJ...356..359H}
\begin{equation}
\label{EqHernquist}
\Phi_{\rm{bulge}}(r)=-\frac{\rm{GM_{{bulge}}}}{r_{\rm{bulge}}+r}\mbox{.}
\end{equation}
This system is also influenced by the MW (as described in the Discussion of Paper I) so we keep the added potential combination that represents the MW in the previous model. 

The numerical simulation for this paper will focus on a simulated set of satellite galaxies around M31. Following the model of Paper I, a plane with equation $z = 0$ and containing 30 satellite galaxies is created - here the $z$ axis points perpendicular to the stellar disc of the M31 potential. The $z=0$ plane can be rotated about the $x$ axis to explore the variations of orientation with regard to the major halo axes. Hence the $z=0$ plane is considered $\theta=0$, and as the plane is rotated about the $x$ axis (where $\theta$ increases), it approaches the $y = 0$ plane. The satellites are spread at a radial distance between 50 and 250\ \kpc\ and with a scatter of $\pm$\ 5\kpc\ perpendicular to the satellite plane. The orbits of the satellites are restricted to elliptical orbits (`$\epsilon$' = 0.5 -1.0) on the initial plane at t=0, by giving 0$\kms$ velocity to the velocity component perpendicular to the plane. 

\begin{figure*} 
\centering
  \subfigure[$q$=1.0]{\includegraphics[width=58mm]{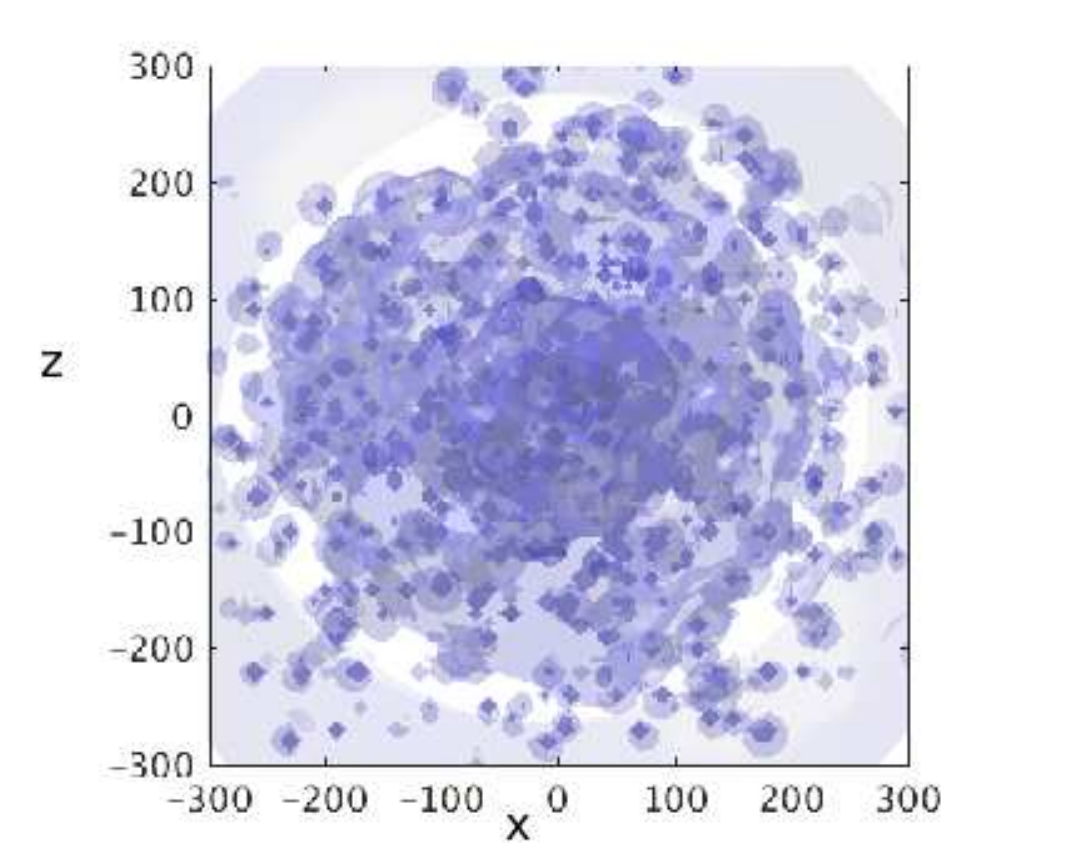}}
  \subfigure[$q$=0.8]{\includegraphics[width=58mm]{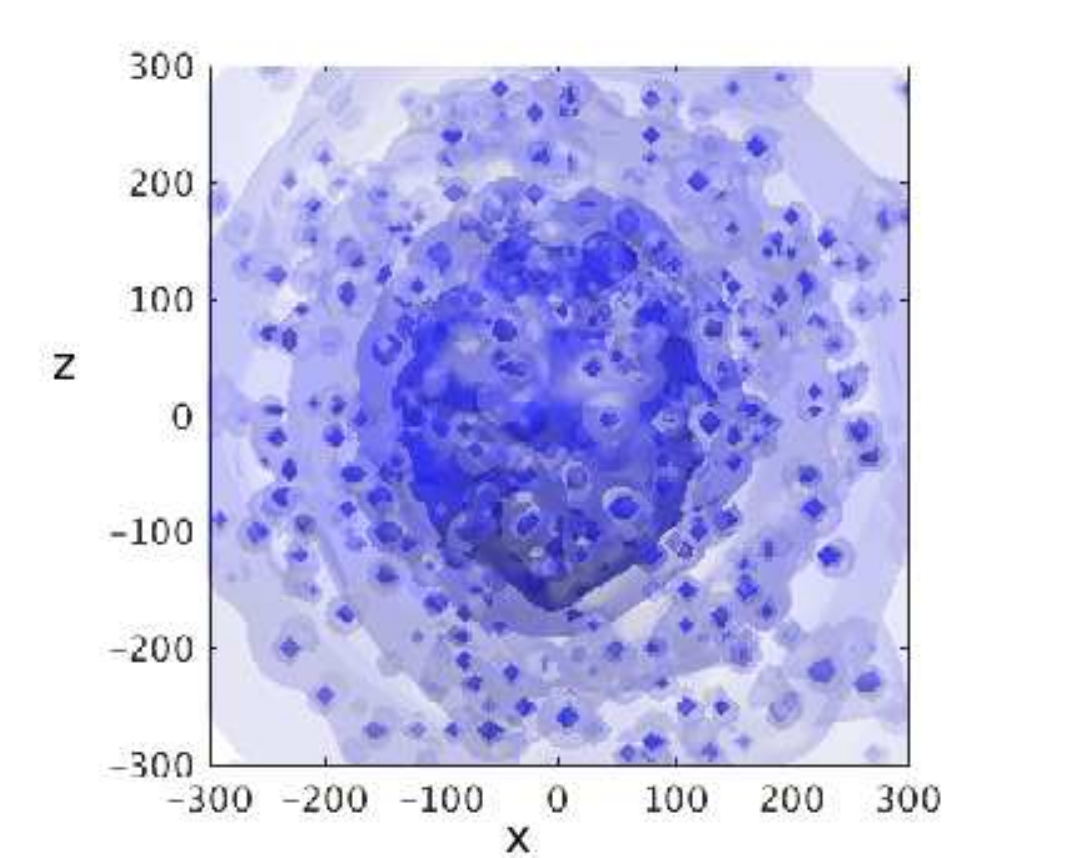}}
  \subfigure[$q$=1.67]{\includegraphics[width=58mm]{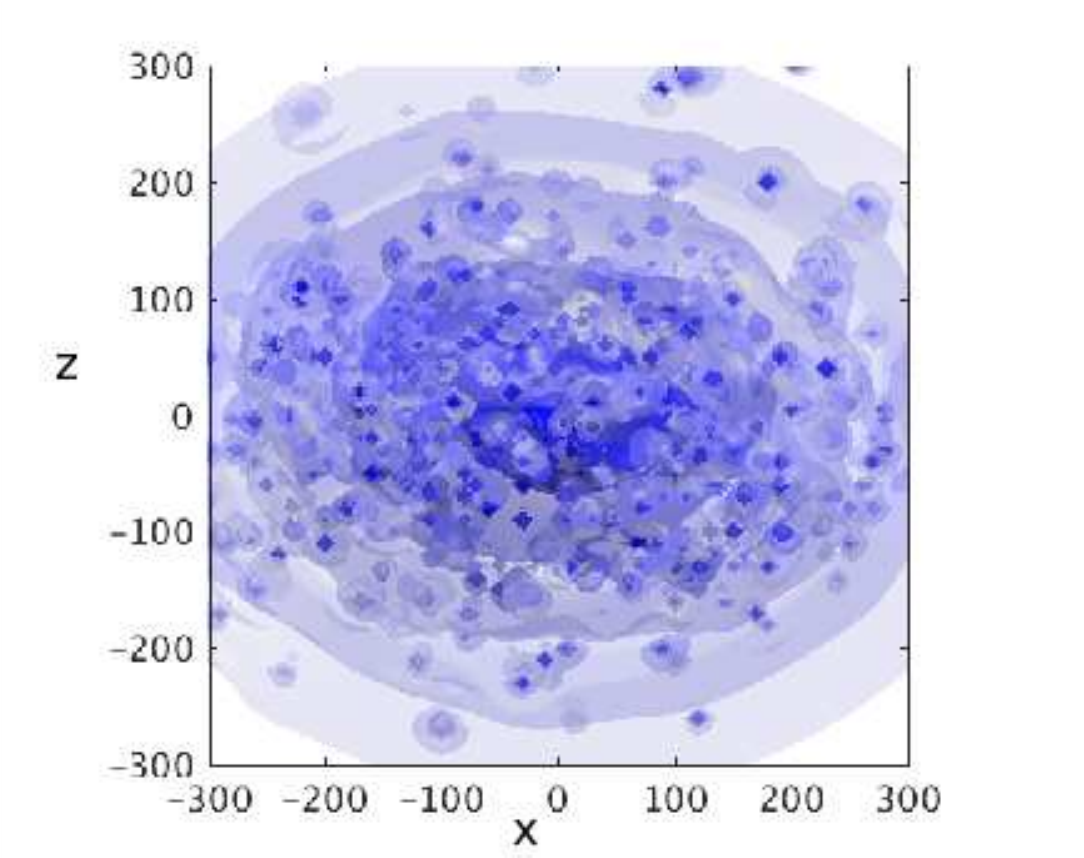}}
  \caption{Host halo shape with 100 subhalos- mean subhalo mass of  $=10^9$ $\msun$ - varying diffuse halo flatness ($q$)}
\label{Fig.2}
\end{figure*}

\begin{figure*}%
\centering
\includegraphics[width=1.0\textwidth]{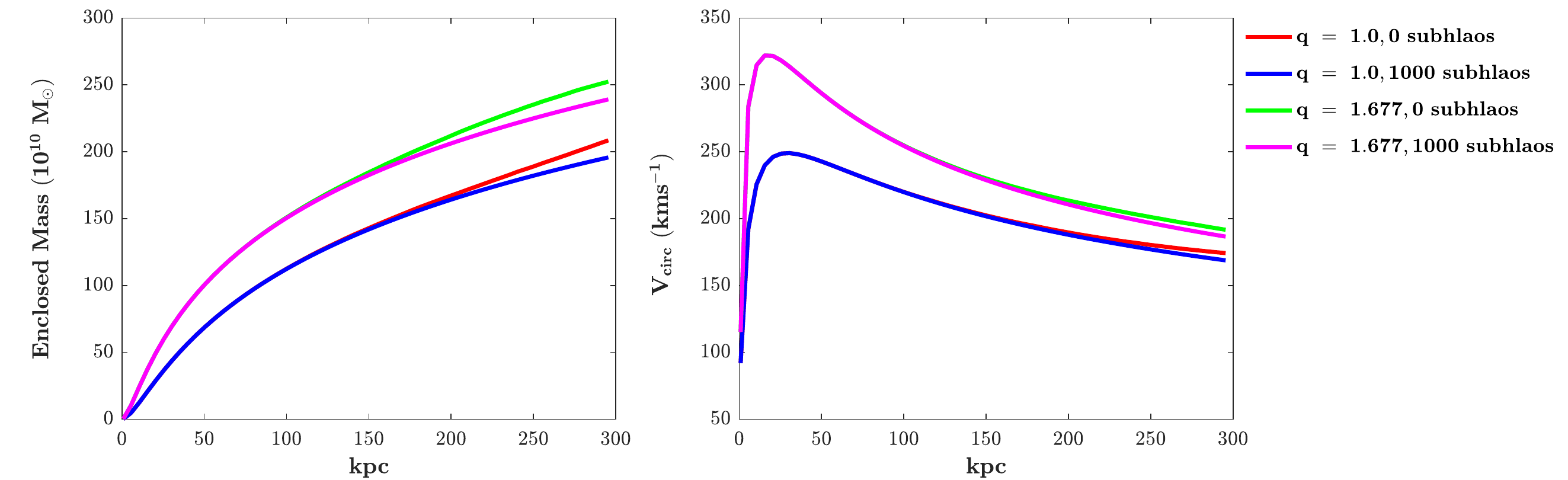}
\caption{Enclosed Mass, and $V_{circ}$ in a dark matter halo including 1000 $=10^8$ $\msun$ subhalos - varying diffuse halo flatness ($q$)}
\label{Fig.2a} 
\end{figure*}

As mentioned in the Introduction, our galactic neighbourhood of M31 and the MW is predicted to play host to hundreds of satellite galaxies and globular clusters that are yet to be detected. To test the dynamic effects of subhalo mass, number and distribution in a plane of satellite galaxies, we add dark subhalos in the manner described below. Subhalos are introduced to the surroundings by placing point mass particles, where radial distances $\geq$50\ \kpc\ are chosen from a Gaussian distribution with a mean of 50$\kpc$\ and standard deviation 250$\kpc$\ from the center of the M31 potential. The minimum distance for subhalo placement is selected at 50$\kpc$, as those too close to the host halo are unlikely to survive. The number and mass of subhalos are varied to observe the effects of these parameters on the satellite plane, but the total mass of the dark matter in the virial radius (250$\kpc$)- $\sim$ $1.03\times10^{12}$$\msun$ for M31- is kept constant through all runs. As shown in Fig.~\ref{Fig.2} and ~\ref{Fig.2a} , the distribution of subhalos follows the shape of the M31 dark matter halo. To ensure the environment of dark matter maintains the given flattening (without a dramatic change in shape), the initial positions and velocities for the tests are taken after the subhalo population has  been integrated forward over 10 Gyrs in a M31-only simulation. Masses for each subhalo are determined by two different methods; 1) a uniform value for all subhalos; 2) chosen from a normal distribution of values where the mean has a value of either $10^9$, $10^8$ or $10^6$ $\msun$ (after Sec.~\ref{Stb_MD}). 

Taking the numerical model described above, we integrate the system forward for 5 Gyrs and use snapshots taken at 0.1 Gyr intervals to calculate how many satellites still remain within $D_{rms}=$ 15 \kpc\ of the initially set plane, and to find the best fit plane to the distribution of satellite galaxies at each timestep (using the method described in Paper I). If we cannot detect 10 or more satellites on a plane, we consider that timestep as a time where no planar structure can be observed. 

\section {An Environment of Dark Halos}\label{Stb_DH}
 
The tests in Paper I show that planes with satellite masses up to $10^{10} \msun$ survive for 5.0 Gyrs if satellites had no initial perpendicular velocities with respect to the plane. But we know that these host halos are not completely devoid of other structures. We first ran preliminary tests to observe what effects 0, 50, 100, and 500 dark subhalos of mass $10^9\msun$ have on a plane of satellite galaxies of $z=0$ residing in a spherical host halo. 

\begin{figure*}%
\centering
\includegraphics[width=1.0\textwidth]{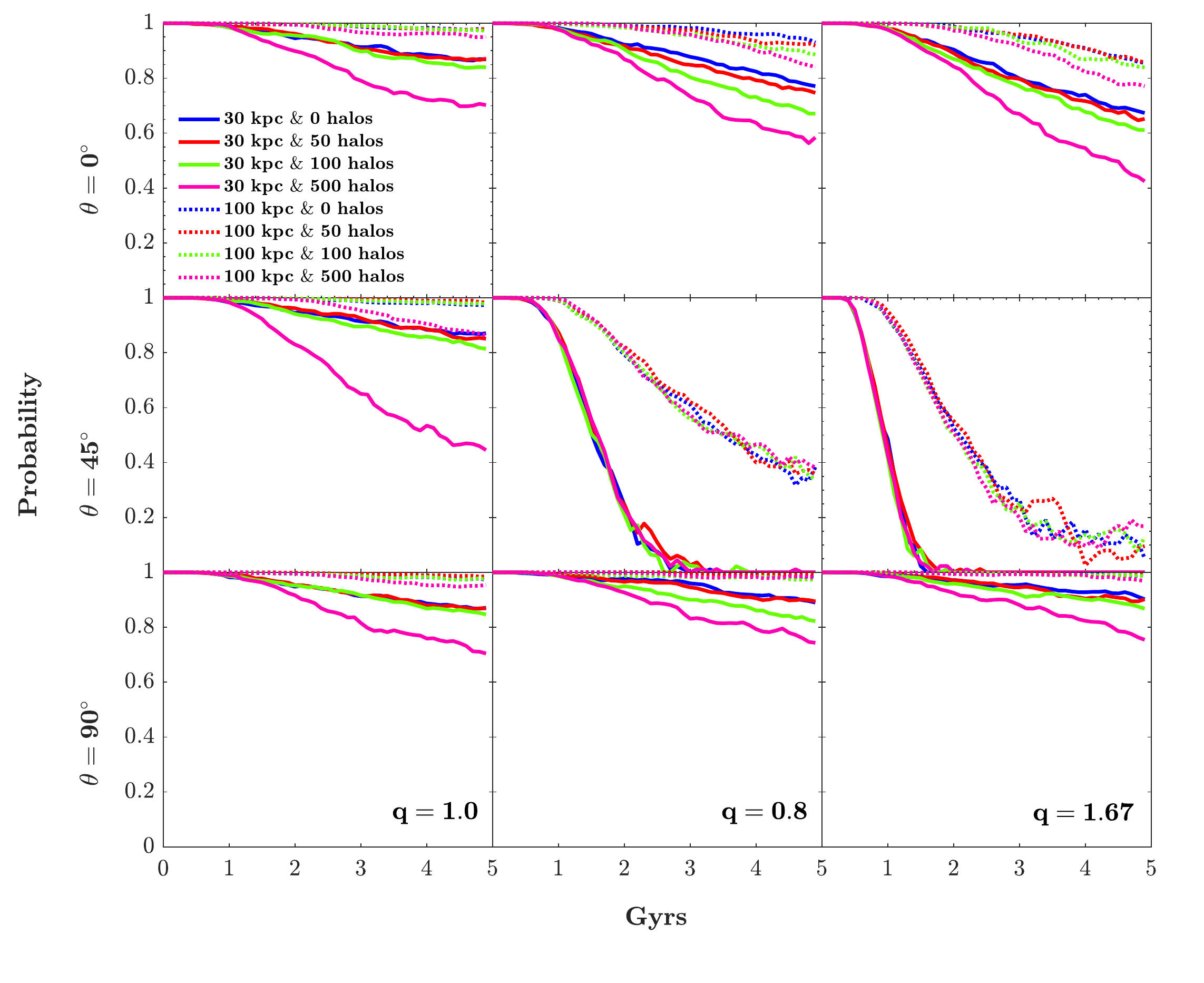}%
\caption{Probability of seeing a satellite on a plane initially set at an incline to the $x$ axis with mean subhalo mass of  $=10^9$ $\msun$ - varying diffuse halo flatness ($q$) and number of subhalos}
\label{Fig.1} 
\end{figure*}

The results, shown in Fig.~\ref{Fig.1} confirm the idea that a shorter lived plane could be the outcome of a distribution of subhalos even for $\theta$=0$\degree$. Being surrounded by 100 halos brings the probability for a satellite to remain on the initial plane after 5 Gyrs to $P=0.8$, which is barely a $5\%$ decrease. However the 500 halos create enough disturbances to the initial plane of satellites to show a clear difference to the other runs. Even though the decline of the probability value happens gradually, by 5 Gyrs the probabilities attain $P$ $\sim$ 0.6. The increased number of halos results in gravitational interactions that are sufficient to alter orbits out of the planar formation. At the end of 5 Gyrs, we can expect around 60\% of the satellites to be seen on the original plane in a spherical halo. Planes of 30\kpc\ thickness have less chance of surviving in flattened dark halos, with $\theta$=45$\degree$ planes not detectable in their initial position after 3 Gyrs. Although we can separate the effects of varying subhalo numbers for stable planes inclined at $\theta$=0$\degree$ in flatter halos, the effects of misalignment will disperse the satellites too quickly to clearly distinguish between subhalo number effects. We can, however, find planes of more than 100\kpc\ thickness up till the end of the run, even though for the most extreme flatness values ($q$=1.677) final $P$ values describe a more random distribution of satellites. 

In conjunction with the results in Paper I, we note that the largest decrease in probability is seen in planes formed with the greatest misalignment to the host halo major axis, $\theta$ = 45$\degree$. Equatorial and polar alignments ($\theta$ = 0$\degree$ and 90$\degree$) show nearly identical results in Fig.~\ref{Fig.1}, as well as Paper I. Therefore, for the purpose of clarity, we plot only the probability ($P$) of seeing a satellite on an initial plane of $\theta$ = 0$\degree$\ and 45$\degree$\ (inclined to the $x$-axis). We will also explore the average number of interactions for a satellite on a plane and the evolution of a plane's thickness for the integration period.

\subsection{A Mass Distribution of Dark Halos}\label{Stb_MD}

As the subhalos' effects on the planes are purely due to gravitational interactions in our tests, it is valid to assume that the disturbance on the plane will be affected by the mass of the subhalos. We conduct the same test of numbers for a mean subhalo mass of $10^8$ $\msun$, but with the same number of subhalos. The smaller mass subhalos have less influence on the plane of satellites, keeping the probability of finding the initial planes above $P\leq$0.85 for entire the 5 Gyr period, thus leaving all tested planes relatively unperturbed. Disruption to a stable plane of satellites depends on the number of subhalos, but it's comparable to their masses -i.e. the notable disturbance that is imparted by $\sim$100 halos of $10^{9}\msun$ average mass, is made by $\sim$1000 subhalos of $10^{8}\msun$ average mass.                

However, by including additional dark matter subhalos in our simulations, we are adding more dark matter into the environment. In order to keep our total dark matter mass the same as before, we decrease the dark matter mass that is in the diffuse halo, with the decreased mass equalling the mass that is introduced  by the subhalos as detailed in Eq.~\ref{EqH} and shown in Fig.~\ref{Fig.2}. As described in Sec.~\ref{sec:NumMod} and shown in Fig.~\ref{Fig.2a}, the total diffused dark matter distribution of M31 without a subhalo population, closely resembles the diffuse halo + subhalo combinations used in the tests. We test how a collection of subhalos with a mean mass of $10^6$, $10^8$, and $10^9$ $\msun$\ can change a plane of satellites. \citet{2009MNRAS.398.1150B}\ claim that around 10\% of total host halo mass is contributed by subhalos, so we shall be testing halo mass fractions that range from 5 to 50\%.

\begin{figure*}%
\centering
  \includegraphics[width=1.0\textwidth]{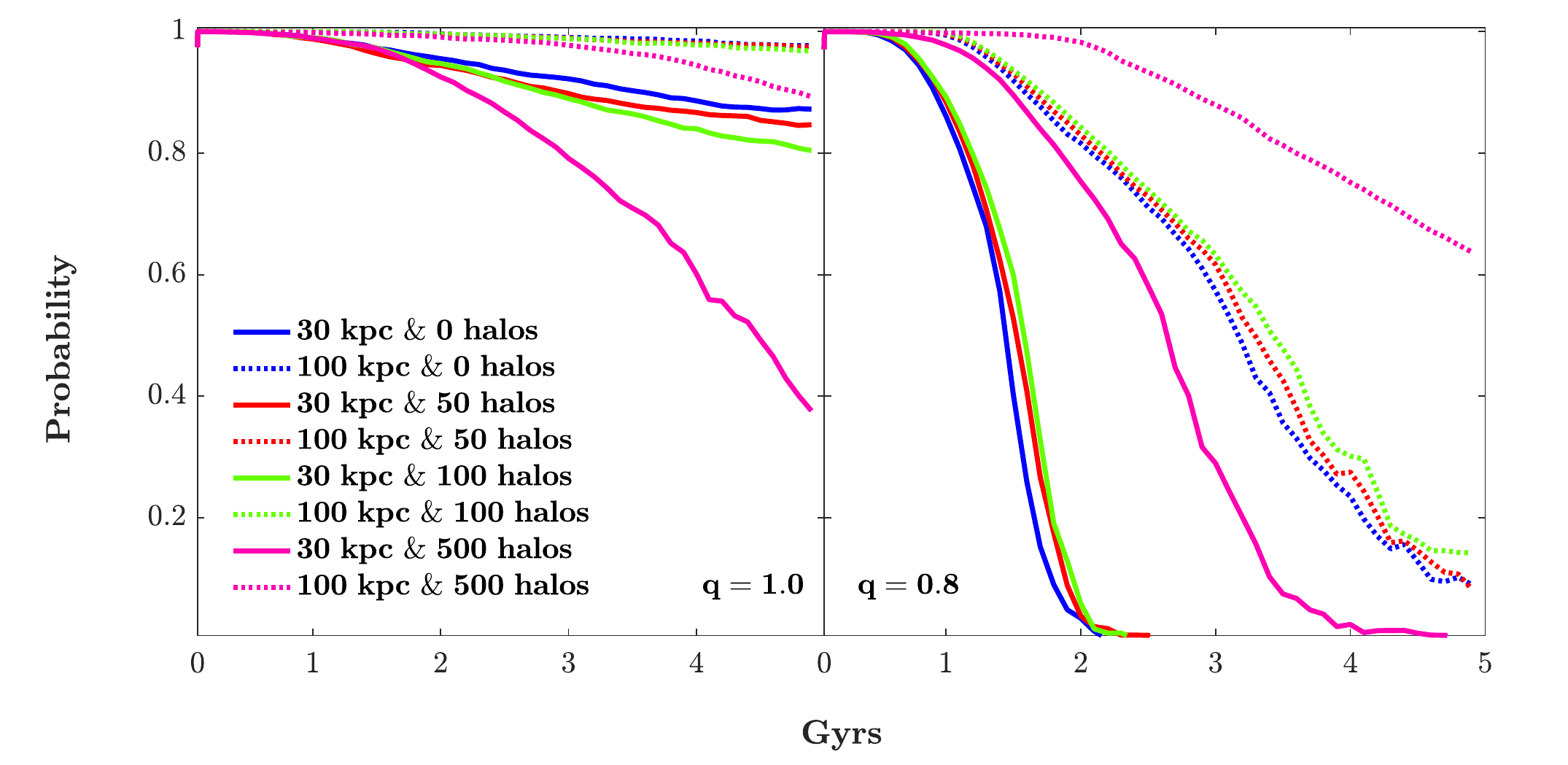}
  \caption{Probability of seeing a satellite on the initial plane with mean subhalo mass of  $=10^9$ $\msun$ - varying diffuse halo flatness ($q$) and number of subhalos (using a plane inclination of $\theta$ = 45$\degree$)}
\label{Fig.3}
\end{figure*}

\begin{figure*}%
\centering
  \includegraphics[width=1.0\textwidth]{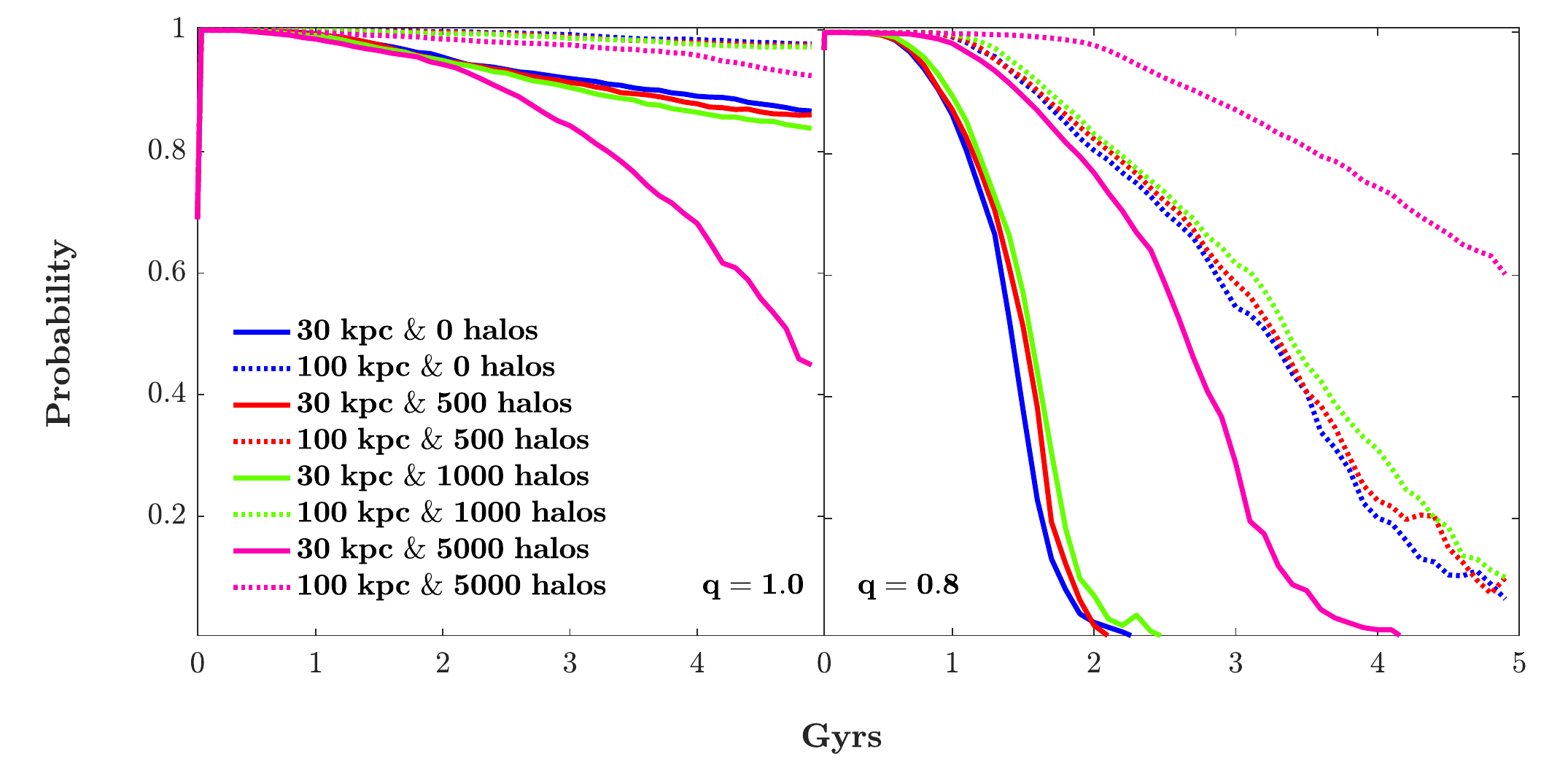}
  \caption{Probability of seeing a satellite on the initial plane with mean subhalo mass of  $=10^8$ $\msun$ - varying diffuse halo flatness ($q$) and number of subhalos (using a plane inclination of $\theta = 45 \degree$)} 
\label{Fig.4}
\end{figure*}

The panels of Fig.~\ref{Fig.3} present the results when this modification is done. They also present the stability of planes in non-spherical halos, for a flatness of $q$\ = 0.8 (Eq.~\ref{EqFl}). Results shown in  Fig.~\ref{Fig.4}\ are similar to those in Fig. ~\ref{Fig.3}, where the drops in probability and the rate of the decline show comparable characteristics to each other. The `blue' line indicating situations where no subhalos are present in the surroundings provides a baseline for the effects created by the halo shape and plane alignment. In comparison, other lines denoting 50, 100, 500 and up to 5000 halos (even though of varying masses) are positioned close to the blue zero line. The number of subhalos changes the probability significantly only after crossing 500 (for $10^9$$\msun$\ mass) or 5000 (for $10^8$$\msun$\ mass). As the subhalo numbers increase, a higher percentage of the total dark matter mass is shifted to the subhalo distribution. The varying number of subhalos between Fig.~\ref{Fig.3} and Fig.~\ref{Fig.4} does not alter the observed decline in $P$ that is caused by the non-spherical nature of the diffuse halo in a subhalo-less environment. An interesting observation is the increase in probability shown in the 45$\degree$ inclined planes with the increasing addition of dark subhalos (Fig.~\ref{Fig.3} and Fig.~\ref{Fig.4}). As a significantly lower percentage (0, 5\% and 10\%) of the total dark halos mass is distributed among subhalos, this increase is difficult to distinguish from one set of probability to another. This gets more prominent as the large halo mass fractions ($h_{m}$) are tested, and can be attributed to an alteration of the total dark matter mass distribution, generated by subhalos.

As in Fig.~\ref{Fig.1}, when the host galaxy's dark halo is a spherical NFW potential ($q$ = 1.0), the planes are likely to exist throughout the 5 Gyrs run, despite having dark subhalos scattered in the surroundings. For a mean subhalo mass of $10^9\msun$, an average number of $\sim$ 100 is too small to make an impact on the plane of satellites. Although the number of subhalos increases as we move to smaller masses, their gravitational effects are negligible, even with $\sim$500 subhalos of $10^7\msun$. We compare results of the most unstable planes in Fig.~\ref{Fig.5} to detect any variations due to flatness of the diffuse halo and/or the subhalo mass fraction. For $q$\ = 0.8 and 1.67 the flattening of the halo in combination with the reduction of dark halo mass and the subhalo environment takes a heavy toll on the plane and is likely to dissipate the plane of satellites by 3.5 Gyrs. For an extreme flatness value ($q$=0.6) the probability of seeing a satellite on a plane is reduced to nearly 0 by 2.5 Gyrs. The reduction of $P$ for smaller $q$ (or larger flattening) is expected as it follows the findings of Paper I, but the effect is augmented by subhalo interactions and the diffuse  mass from the flattened halo. Even more significantly, the changes seen for each number of halo flatness in Fig.~\ref{Fig.5}, are very small, if not negligible, and the plots show a remarkable likeness to each other for the different mass values. This suggests that the reduction in the `diffuse' dark halo mass and flatness is more significant than the mean mass and number of subhalos. 

\subsection{Varying Dark Subhalo Mass Fraction}\label{Stb_MF}
    
The fraction of the total dark matter mass included in subhalos (subhalo mass fraction) is shown to play a larger role in creating surroundings that are advantageous and disadvantageous to long lived satellite galaxy planes. To investigate the role of this property in more detail, we conducted further tests where the subhalo mass fraction is varied as 0.1, 0.2, and 0.4. The number of subhalos in each test will vary according to the fraction of the total dark matter mass that is represented in subhalos {i.e. to get a subhalo mass fraction of 0.1, one would have $\sim$100 subhalos with a mean mass of $10^9$$\msun$, or $\sim$1000 subhalos with a mean mass of $10^8$$\msun$ and $\sim$100000 subhalos with a mean mass of $10^6$$\msun$}. In the set of tests shown in Fig.~\ref{Fig.1} and afterwards, where the fraction of subhalo is 10\% of the M31 dark halo mass, halo mass fraction ($h_{m}$) will be 0.9 for Eq.~\ref{EqH}). We will add these variation parameters to the model used above. 

\begin{figure*}%
\centering
  \includegraphics[width=1.0\textwidth]{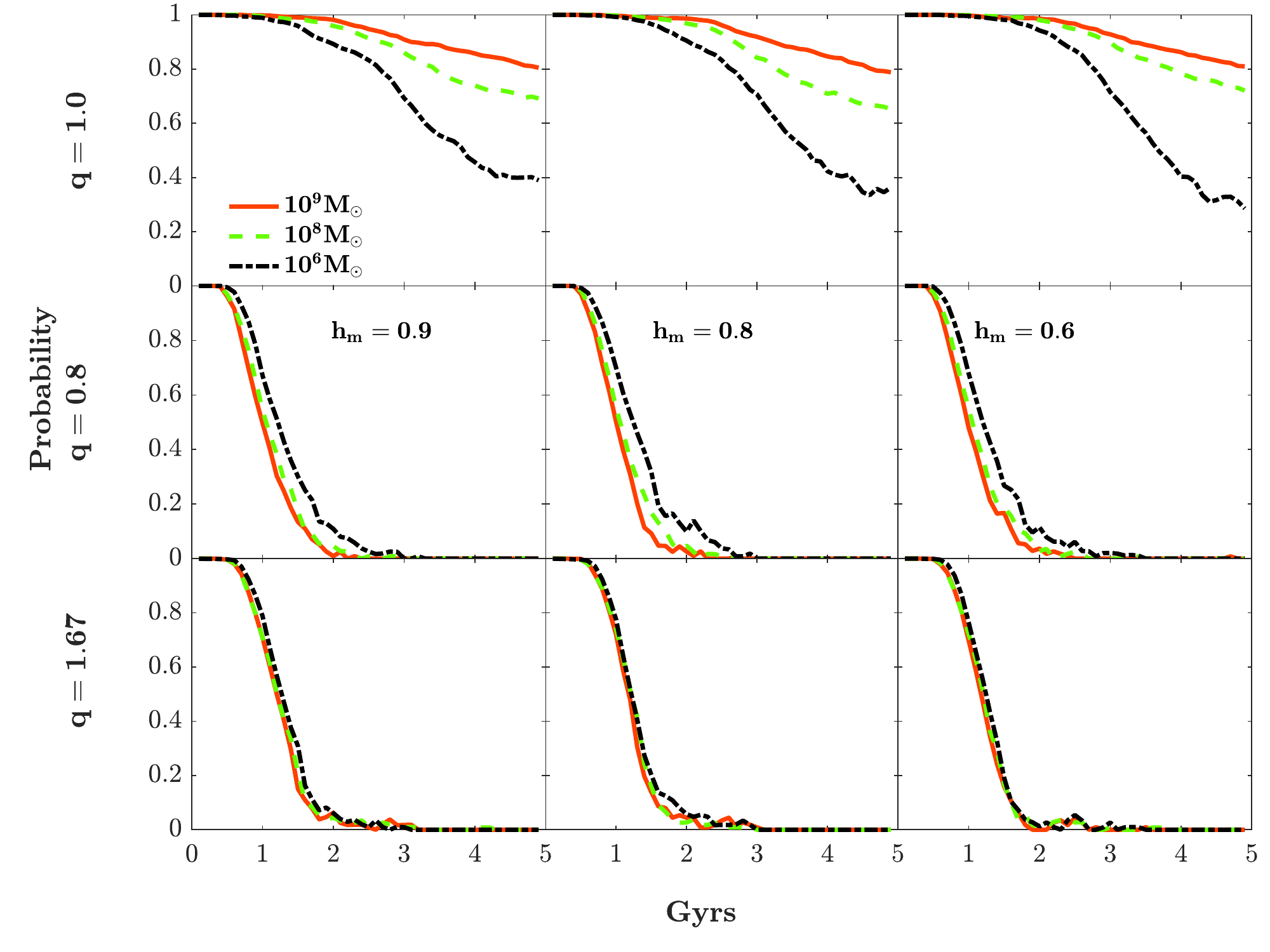}
  \caption{Probability of seeing a satellite on the initial plane (30$\kpc$)- varying halo mass fraction ($h_{m}$),diffuse halo flatness ($q$), and mean mass of subhalos (using a plane inclination of $\theta = 45 \degree$)}
\label{Fig.5}
\end{figure*}

The resulting probability plots for satellite galaxy planes initially set at $\theta$ = 45$\degree$ in halos of varying flatness is shown in Fig.~\ref{Fig.5}- as 45$\degree$ planes experience the greatest effects from plane misalignments with the dark halo axes. The variations of mean subhalo mass follows Sect.~\ref{Stb_MD}. Several significant observations can be made when comparing this set of figures (Fig.~\ref{Fig.5}) with Fig.~\ref{Fig.1}. For all shapes of host dark halos, the change across various mean subhalo masses show little difference for one subhalo mass fraction. A dark halo where 20\% of the mass is contained in dark halos will reduce a satellite galaxy's probability of being seen on it's initial plane by $\sim$10\% if the main halo is spherical in shape, and increase the probability by $\geq$ 5\% by the end of 5 Gyrs for a slightly flat halo ($q$=0.8). In dark halos of more extreme flatness ($q$=0.6/1.67 for oblate/prolate halos) the incline of the plane in the oblate halo is enough to disrupt the plane by 2.5- 3.0 Gyrs in Fig.~\ref{Fig.5}, while the number of subhalos and their masses make little if no difference in the rate of the decline of the probability for flattened halos. 
		
To further test the influence of the number of subhalos vs. subhalo mass fraction, we altered the tests to contain a fixed number of subhalos, and the remaining mass to be held in the diffuse dark halo. Resulting figures Fig.~\ref{Fig.3} and ~\ref{Fig.4} show very little difference from each other, though the number of subhalos has increased by an order of magnitude. The dark halo holds $\sim$0\%, 5\%, 10\% and 50\% of the dark matter mass in Fig.~\ref{Fig.3} for the subhalo numbers 0, 50, 100 and 500 respectively where average subhalo mass is $10^{9}$$\msun$. The same fractions apply to Fig.~\ref{Fig.4} where subhalo numbers vary from 0, 500, 1000 to 5000 and average subhalo mass is $10^{8}$$\msun$. The variation from Fig.~\ref{Fig.5} takes place for high values of subhalos mass fraction ($\sim$50\%). For spherical dark matter distributions, $P$ drops to 0.3 at 5 Gyrs, and only a third of the satellite population is likely to be seen on the plane they began on. This is more than a 50\% drop in probability from when only 10 and 20 \% of the dark matter mass was in the subhalos. This trend is replicated in the $10^{8}$$\msun$ subhalo distributions, where the larger subhalo numbers do not make a visible change in the two plots. 

\begin{figure*}
\centering
\includegraphics[width=1.0\columnwidth]{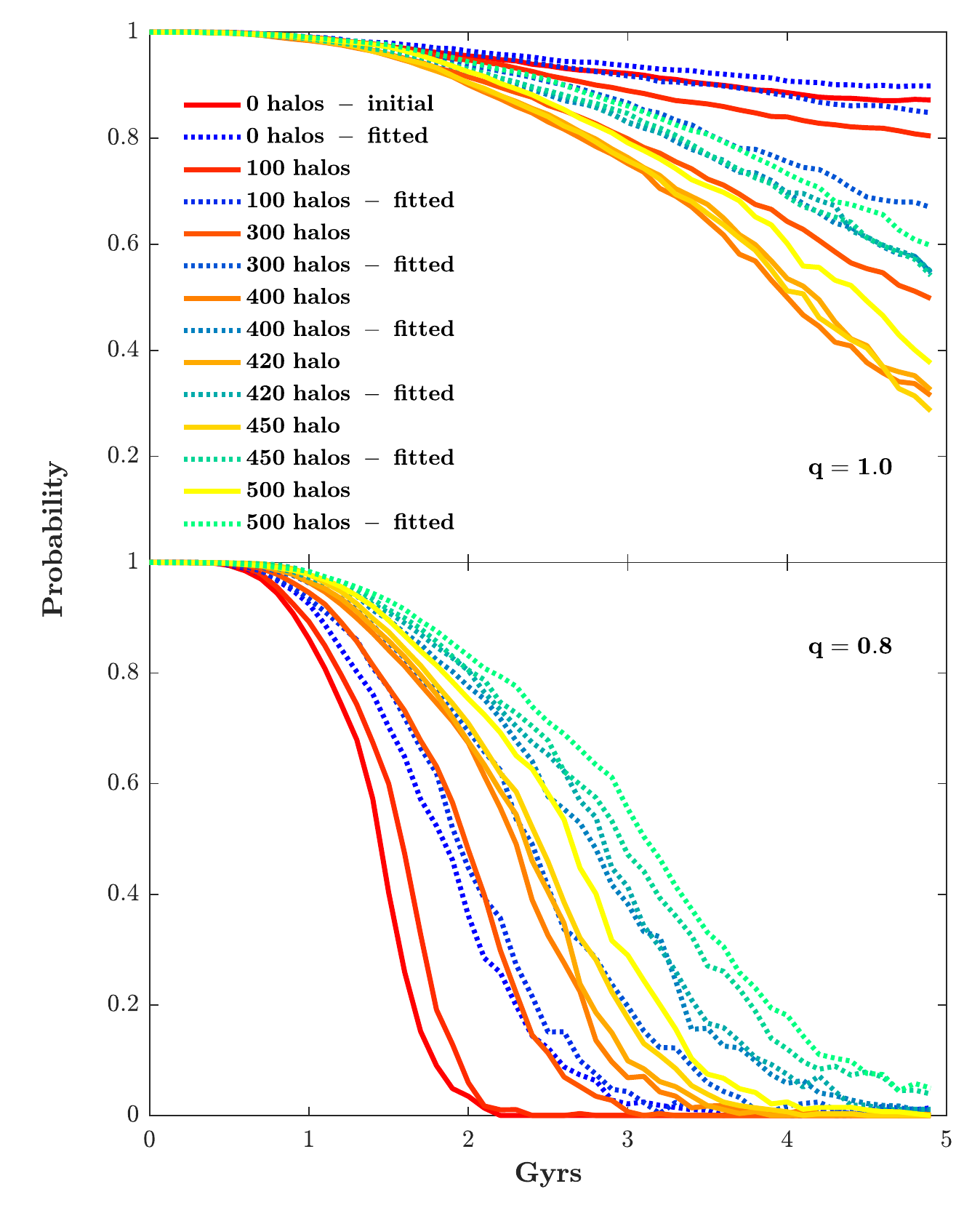}	 
\caption{Probability of seeing satellite on initial plane, mean subhalo mass of  $=10^9$ $\msun$ - halo flatness $q$=0.8 ($\theta$ = 45 $\degree$)}
\label{Fig.6}
\end{figure*}

The most substantial divergences are seen in the misaligned planes ($\theta$=45$\degree$) where the 500 and 5000 subhalo number models ($h_m$=50\%) present a larger probability of survival than smaller numbers. To test where this curious (and almost counter-intuitive) behaviour starts, we explored plane longevity for intermediate subhalo mass fractions for spherical and flat halos. Fig.~\ref{Fig.6} showcases results from our integrations for an average subhalo mass of $10^9$$\msun$. It is apparent that although all tested subhalo numbers lead to a scattered distribution of satellites, the rate of decline largely varies with subhalo numbers that correspond to mass fractions $\leq$ 30\%. Starting from 300 subhalos, for each 100 additional subhalos included, the time taken to reach $P$=0.0 is extended by $\sim$ 0.5 Gyrs. When the probability for the best-fit plane (of 30\kpc\ thickness) is calculated, we note that the satellite distribution shows some planar formation for $\geq$ 1 Gyrs if the orientation is not restricted to the initial plane. However, by 5 Gyrs, we are extremely unlikely to detect the satellites on a plane when it is misaligned with the major axes of a non-spherical halo, regardless of a difference in the subhalo population. Similar results were seen in tests conducted for $10^{8}$$\msun$ subhalo mass averages.       

However, all models show that planes are disrupted as dark subhalos are included. This disruption can double the thickness of the plane by 5 Gyrs, for the most stable models (spherical dark halos with small numbers of subhalos). Fig.~\ref{Fig.7} presents the effects of halo flatness on the width of a misaligned plane. Plane width is calculated using the difference between the perpendicular distances of the furthermost members on each side of the the best-fit plane at each timestep. Planes start with initial thickness $\leq$ 10$\kpc$\ and extend up to 80 $\kpc$. The 14 $\kpc$ thickness shown by the VTPoS is reached during the 1-2 Gyr period of the satellites' orbits. 

As it is evident that increased subhalo numbers do not always lead to decreased longevity and stability, we examined how subhalos interact with satellites. Fig.~\ref{Fig.8} presents the number of subhalos that come within 10$\kpc$\ for an average satellite galaxy set on a plane, given varying subhalo numbers and masses. The increase in subhalo numbers clearly lead to more `close' encounters with subhalos, and the average interaction numbers do not vary from the angle of incline. Recorded encounters increased by a factor of 10 for a subhalo number change from 100 to 500 (for $10^{9}$$\msun$). This factor of 10 increase is also reflected in the smaller mass subhalos.       

\section{Discussion}\label{sec:Disc}

From the findings of Paper I and \citet{Bowden2013}, we know that the stability of satellite galaxy planes is affected by the host halo shape and plane orientation with regard to the axes of the host halo. $\Lambda$CDM dark matter only simulations predict dark subhalos, of a range of masses, in the environment of these planes. In our tests, the Gaussian dark substructure distribution follows the general flattening shape of the dark halo. Subhalos can range from $10^{4}$ to $10^{10}$$\msun$ in our simulations and subhalo numbers depend on the fraction of total dark matter mass shifted to subhalos and their mean mass. 

Planes in a smooth halo behave as predicted in Paper I, with all alignments to the halo axes reflecting the same overall probabilities during the integration period. Adding a number of subhalos into the domain notably changes the probability of seeing a satellite on its initial plane only when $\sim$40\% of the entire dark matter mass is contained in the subhalo population, regardless of the number and mass of subhalos that it may contain. The number of dark halos predicted by $\Lambda$CDM simulations shows slight variation from one to another, and M31 like hosts are likely to be surrounded by approximately 0-10, 50-200 and $\geq$ 500 subhalos with respective individual masses of $\sim$$10^9$, $10^8$ and $10^7$\ $\msun$ \citep{2009MNRAS.398.1150B, Sawala2016}. They also predict larger numbers of smaller subhalos ($\leq 10^{6}\msun$), but in total the average contribution of the subhalo distribution to the overall dark matter mass is often $\leq$ $20\%$ \citep{2009MNRAS.398.1150B}. Comparing just these subhalo mass fractions with our results (Fig. ~\ref{Fig.1}), such satellite planes are unlikely to be significantly disturbed within 5 Gyrs. The current emergence of Warm Dark Matter models predicts a lower number of subhalos \citep{2017MNRAS.464.4520B}, especially in the smaller scales. Our results above show that satellite planes can maintain a long lifespan ($\sim$ 5 Gyrs) in environments that are both rich and sparse in sub- halos, if the conditions presented in Paper I are met (planar alignment with host halo axes or spherical diffuse halos and containing 40\% of the total dark matter mass in subhalos). 
 
Results from our simulations also vary due to the shape of the dark matter halos as well. The M31-sized dark matter  host halos of dark matter only simulations \citep{2008MNRAS.391.1940M} present a triaxial nature with flatness varying from 0.8 to 0.5, or lower in some instances. With $q$ = 0.8 being the lowest tested flatness in this paper, our results can be treated as an upper limit to the probabilities of seeing a satellite on similar non-spherical distributions of dark matter in the halo. Fig.~\ref{Fig.5} presents the probability of seeing a given satellite on its initial plane of $\theta$=45$\degree$, with respect to halos of spherical, oblate and prolate nature. It is clear that the gravitational effects of the large halo cause these planes to dissipate. 

Recent work on the shapes of dark halos of large galaxies present $a/c$ axis ratios averaging from 0.7- 0.4 \citep{2008MNRAS.391.1940M}. The NIHAO Hydro simulations\citep{2015MNRAS.453.1371M} measure triaxial halos for smaller halos ($\sim$$10^{11}$$\msun$) and more spherical large host halos ($\sim$$10^{12}$$\msun$) of major axis ratios averaging at 0.8- which matches the \citet{Ibata2001} halo-shape measurements. Our results in Fig.~\ref{Fig.5} show that even a small flatness like $q$=0.8 can create significant disruption for misaligned planes of satellites in 5 Gyrs. As seen in Figs.~\ref{Fig.1},~\ref{Fig.3} and ~\ref{Fig.4}, there is very little probability of finding any satellite on its initial plane which is $\leq$ 100\kpc\ in thickness, by the end of a 5 Gyrs integration period for a halo flatness of similar values (which is over seven times the calculated thickness of the VTPoS). 
 
As a larger fraction of the total dark matter mass is transferred into subhalos, it becomes difficult for the overall dark matter distribution to hold its initial shape. The flatness of the dark matter decreases ($q$ heads towards 1.0) as the subhalos settle into the most stable formation, and redistributes the dark matter mass, now almost dominated by subhalos. This creates a more spherical environment that allows satellite planes to last longer in their original plane. Even the most misaligned planes take 2 Gyrs longer to reach $P$=0.0 (for 30 \kpc\ thickness). The probability of finding a plane containing more than 30\% of the total satellite population (10 or more) within a thickness of 100 \kpc\ is relatively high. Although cosmological simulations do not show such a high mass fraction in the subhalo population \citep{2009MNRAS.398.1150B}, it is likely that such a situation will lead to constantly changing halo shapes, that will effect a satellite plane's stability throughout a 5 Gyr period of time.  

\begin{figure*}
\centering
\includegraphics[width=1.0\columnwidth]{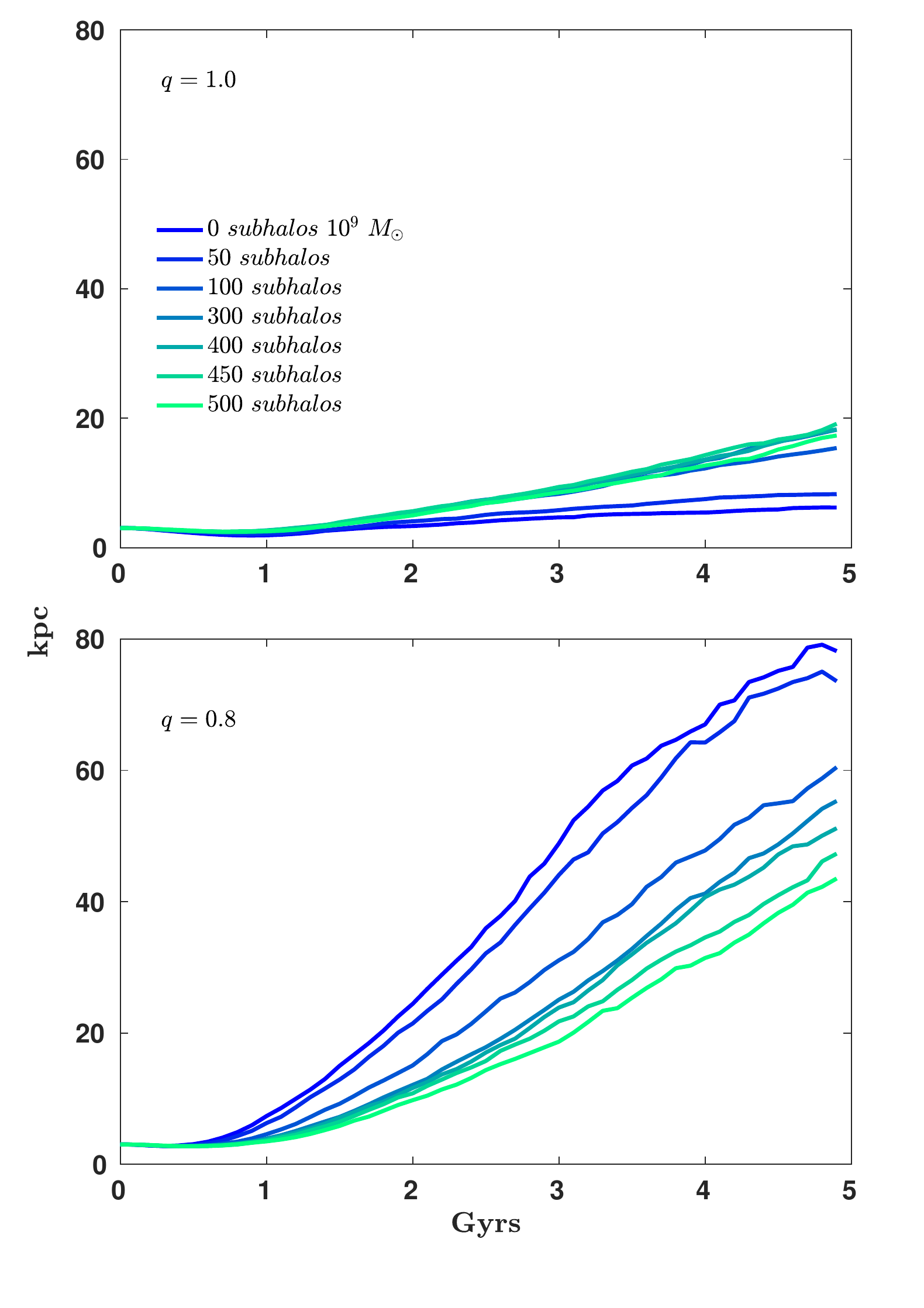}	  
\caption{Thickness of best-fit plane - varying dark subhalos numbers, mean mass $10^9$ $\msun$ }
\label{Fig.7}
\end{figure*}

\begin{figure*}
\centering
\includegraphics[width=1.0\textwidth]{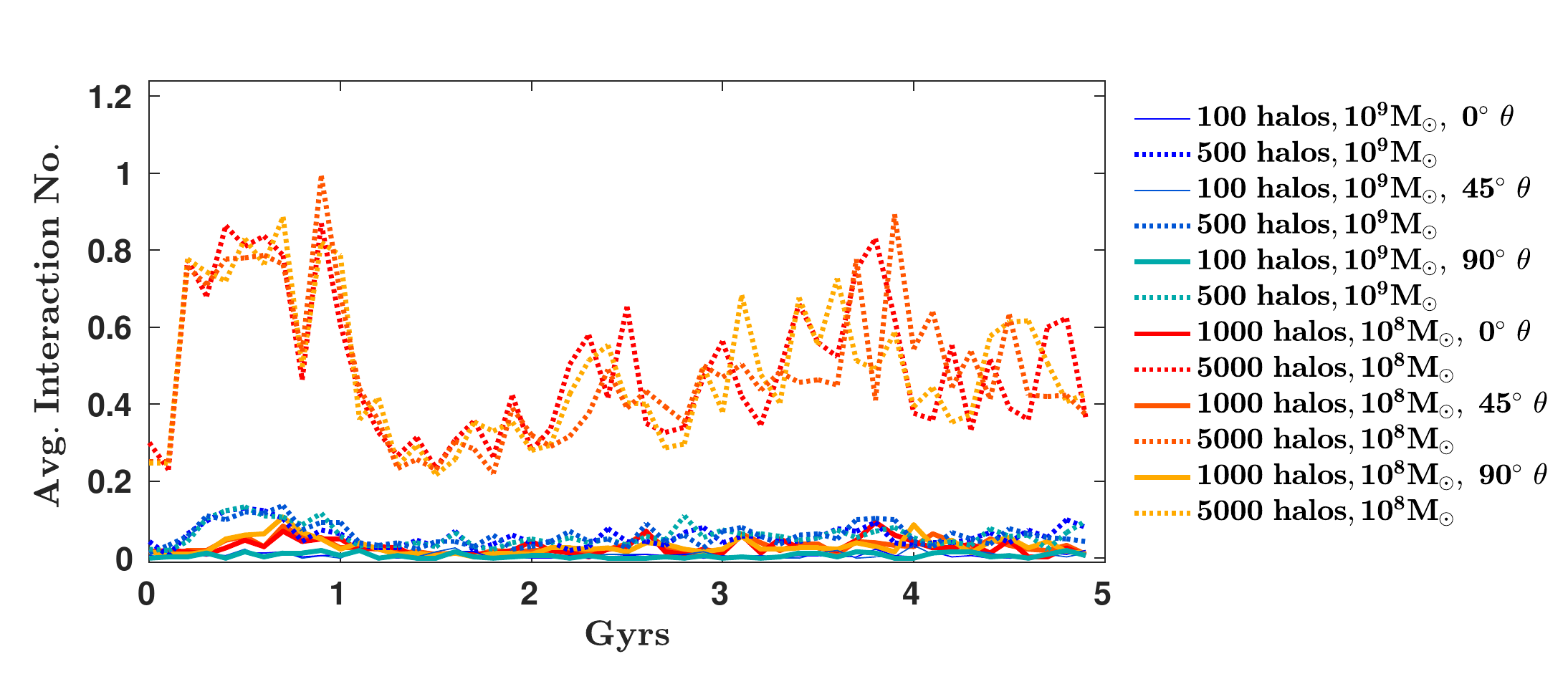}	  
\caption{Average subhalo interaction numbers within 10\kpc- varying dark subhalos numbers and mean subhalo mass }
\label{Fig.8}
\end{figure*}

We corroborate a central result from Paper I: the evolution of orbits of satellites in misaligned planes found in flattened halos leads to the dissolution of the initial plane the satellites were found in. The precession of orbits due to halo misalignment will increase the thickness of the planar formation (Fig.~\ref{Fig.7}), where a 30 \kpc\ plane could no longer be found by the end of 2 Gyrs in the simulation, a lifetime that is short enough to be thought of as a transient or temporary satellite plane. The almost uncanny similarities between the panels in Fig.~\ref{Fig.5} lead us to conclude that the stability of a misaligned planar formation of satellite galaxies in an M31-like host galaxy is unlikely to be determined by the host’s dark subhalo population.

\section{Conclusions}\label{Concln}

Explanations for the stability and longevity of the M31 plane of galaxies have taken various avenues. By modifying a numerical model of an M31 and MW potential, we ran a set of simulations of 30 satellite galaxies with varying environments and properties. We set satellite galaxies on an initial plane to calculate the probabilities of seeing a satellite on their initial or best-fit planes in a simulation time of 5 Gyrs. The following properties were varied both individually and in combination: satellite velocities perpendicular to the plane, mass of satellites, number of dark subhalos, incline of satellite plane to the host halo, flatness ($q$) of the host halo and the mass of host halo replaced by subhalos. From our findings so far, the following conclusions can be inferred by the statistics obtained from our set of simulations.
\begin{enumerate}[(1)]
\item Dark subhalos have a notable effect on a satellite plane. Subhalo interactions contribute to shortened life spans of planes, although an environment of 500 similar mass subhalos is likely to show differences from an empty environment only after a few Gyrs. 
\item Overall dark halo shape and plane alignments will be the dominant factors in determining the stability and longevity of a satellite galaxy plane. Interactions with dark subhalos may cause the initial planes to increase in thickness at an earlier time step, but the effects will not show significant change for smaller numbers of halos.
\item Where a smaller fraction of dark matter is contained in the subhalos, the overall shape of the dark matter halo is more stable. When the subhalo population counts for over 40\% of the total dark mass, it makes halo flatness more susceptible to significant changes due to the dynamics of the subhalo population- changes that are visible in through the 5 Gyr integration period. These changes are also visible in a plane of satellite galaxies, as it becomes more stable to adjust to a halo that is increasing in its spherical nature. 
\end{enumerate}

With regard to M31, the non-planar satellites would also have an influence on the VTPoS, especially when considering interactions with M33, which is calculated to have a greater mass than the average VTPoS member satellite. However, our tests show that the lifespan of the VTPoS would depend on the shape and stability of its dark matter halo, and its alignment to the halo axes. Therefore, planar structures could be used as tracers of the local dark matter distribution, although finer details such as subhalo mass and numbers would only be detectable in the most stable environments (i.e. spherical dark matter halos).   

\section*{Acknowledgments}
NF acknowledges the Dean's International Postgraduate Scholarship of the Faculty of Science, University of Sydney. GFL and CP acknowledge funding through the ARC Discovery Program DP140100198.

\bibliographystyle{mn2e}
\bibliography{Satellite_Planes_Subhalos_Revised1}

\label{lastpage}

\end{document}